\documentclass[apj]{emulateapj}

\newcommand{\etal}{et~al.}
\newcommand{\deltah}{\ensuremath{\delta H / H}}

\begin{document}
\title{Is There Evidence for a Hubble Bubble?  The Nature
 of Type~Ia Supernova Colors and Dust in External Galaxies}
\shorttitle{The Hubble Bubble And SN~Ia Colors}
\shortauthors{Conley \etal}
\author{
 A.~Conley\altaffilmark{1},
 R.~G.~Carlberg\altaffilmark{1},
 J.~Guy\altaffilmark{2},
 D.~A.~Howell\altaffilmark{1},
 S.~Jha\altaffilmark{3},
 A.~G.~Riess\altaffilmark{4,5},
 M.~Sullivan\altaffilmark{1}
}
\altaffiltext{1}{Department of Astronomy and Astrophysics, University
  of Toronto, Toronto, ON M5S 3H4, Canada}
\altaffiltext{2}{LPNHE, CNRS-IN2P3 and University of Paris
  VI and VII, 75005 Paris, France}
\altaffiltext{3}{Kavli Institute for Particle Astrophysics and Cosmology,
 SLAC, Menlo Park, CA 94025, USA}
\altaffiltext{4}{Space Telescope Science Institute, MD 21218, USA}
\altaffiltext{5}{Johns Hopkins University, MD 21218, USA}
\email{ conley@astro.utoronto.ca }

\begin{abstract}
We examine recent evidence from the luminosity-redshift relation of
Type~Ia Supernovae (SNe~Ia) for the $\sim 3\, \sigma$ detection of a ``Hubble
bubble'' -- a departure of the local value of the Hubble constant from
its globally averaged value \citep{Jha:07}.  By comparing the MLCS2k2
fits used in that study to the results from other light-curve fitters
applied to the same data, we demonstrate that this is related to the
interpretation of SN color excesses (after correction for a
light-curve shape-color relation) and the presence of a color gradient
across the local sample.  If the slope of the linear relation
($\beta$) between SN color excess and luminosity is fit empirically,
then the bubble disappears.  If, on the other hand, the color excess
arises purely from Milky-Way like dust, then SN data clearly favors a
Hubble bubble.  We find that SN data give $\beta \simeq 2$,
instead of the $\beta \simeq 4$ one would expect from purely
Milky Way-like dust.  This suggests that either SN intrinsic colors
are more complicated than can be described with a single light-curve
shape parameter, or that dust around SN is unusual.  Disentangling
these possibilities is both a challenge and an opportunity for large-survey
SN~Ia cosmology.

\end{abstract}


\section{INTRODUCTION}
\nobreak
In an analysis of the luminosity distances of Type~Ia supernovae
(SNe~Ia), \citet{Jha:07} (hereafter J07) presented evidence for an
offset between the Hubble constant measured from SNe~Ia with $2500 <
cz < 7400$ km s$^{-1}$ and from those with $7400 < cz < 45000$ km
s$^{-1}$.  Specifically, the more distant SNe are slightly fainter
than one would expect, implying a high local value of $H_0$.  One
natural explanation is a ``Hubble bubble'' -- a local monopole in the
peculiar velocity field, perhaps caused by a local void in the mass
density.  The analysis of J07 was carried out using the most recent
manifestation of the MLCS light-curve fitter, MLCS2k2, and found
$\deltah = 6.5\% \pm 1.8\%$.  The $\Delta m_{15}$ method
\citep{Prieto:06} gives similar results.  In contrast, galaxy-cluster
distances give $\deltah \simeq 1.5\% \pm 2\%$ \citep{Giovanelli:99,
Hudson:04}, which is marginally inconsistent with the SN result ($1.9
\sigma$).  Earlier studies using SNe~Ia can be found in \citet{Kim:97}
and \citet{Zehavi:98}.  A Hubble bubble could have serious
implications for precision SN~Ia cosmology programs.

In this Letter we test the evidence for the Hubble bubble using three
other light-curve analysis packages: SALT \citep{Guy:05}, SALT2
\citep{Guy:07}, and an unpublished package developed for the 3rd year
Supernova Legacy Survey \citep[SNLS,][]{Astier:06} data (SiFTO).  An
analysis of the same data set with these tools does not support a
bubble if SN data are used to derive the relationship between SN color
excess and peak luminosity, but does if this relationship is required to be
that of Milky Way-like dust. Therefore the question of the Hubble
bubble is one of the nature of SN colors.  A similar result was found
independently by \citet{Wang:07}.

\section{LIGHT-CURVE FITTERS AND THE MEANING OF SN COLORS}
\label{sec:fitters}
\nobreak 
A considerable amount of effort has been devoted to the question of
how to best fit SN~Ia light curves.  The goal is to measure the
relative luminosity distance of different SNe, usually after
correcting for the width-luminosity and color-luminosity
relationships.  These approaches differ considerably in implementation
and assumptions, but generally produce very similar results when used
to estimate the cosmological density parameters ($\Omega_m,
\Omega_{\Lambda}$) \citep{WoodVasey:07}.

In this analysis we consider four fitting packages: MLCS2k2, as well
as three packages developed for use with SNLS: SiFTO, SALT, and SALT2.
The latter three are more similar to each other than to MLCS2k2.  The
details of SiFTO will be presented elsewhere (Sullivan \etal , in
preparation); it is broadly similar to SALT except that the color excess
relation is not imposed during the light-curve fit, but rather when
the results of the fit are converted into a distance estimate.

The critical issue for the question of the Hubble bubble relates to
how the different fitters handle SN color excesses.  All of the models
have two parameters: one that describes the light-curve shape, and
one the color, both of which affect the peak luminosity. For MLCS2k2
these are $\Delta$ and $A_V$, for SALT they are $s$ and $c$.  These
have different technical meanings, but essentially describe the same
effects.  MLCS2k2 breaks the color relation into two pieces: a 
time-independent piece\footnote{Technically, the effective value of $R_V$
depends on the SED, so the effects of this term are weakly 
time-dependent.}, parameterized by $A_V$, and a time-dependent piece
determined by the shape of the light curve, $\Delta$.  The time
independent piece is assumed to follow the CCM \citep{CCM:89} dust
extinction law.  In effect, $A_V$ parameterizes the color of the SN
after correction for a shape-color relation.  SALT follows a similar
prescription, with some critical differences. The wavelength behavior
of the time-independent piece (parameterized by $c$) is derived as
part of the SALT training process; the results are shown in
\citet{Guy:07}.  This relation is reasonably similar to the CCM law,
and is arbitrarily normalized so that $c$ is the $B-V$ color at peak
luminosity (plus a small constant) for any $s$; the same is not true
of $U-B$ and $V-R$, which are functions of $s$.  In the following we
refer to the time-independent piece as the color excess relation.

The real difference is how the color excess relation is used to
correct the peak luminosities. MLCS2k2 makes the simplest assumption --
that it is due to dust.  The wavelength behavior of the CCM law is
parameterized by the ratio of selective to total extinction ($R_V$),
and this is also used to convert the color excess into the magnitude
correction.  The other fitters considered here take a different
approach.  One of the steps in the process of turning their results
into a distance estimate is to empirically measure the slope of the
relationship between the color $c$ and the luminosity correction
($\beta$), as well as between the light-curve shape and luminosity
($\alpha$) by minimizing the residuals with respect to the Hubble
line.  Here the model for the predicted peak $B$ magnitude $m_B$ is
$m_B = M_B + 5 \log_{10} d_L - \alpha \left(s-1\right) + \beta c$,
where $M_B$ is the absolute peak magnitude, $d_L$ is the luminosity
distance, and $\alpha$ and $\beta$ are determined from SN data; the
slopes can be cleanly separated from the intercept without any
difficulties.  In this formulation, the shape-color relation is
absorbed into the $\alpha$ coefficient, and therefore the $\beta$
correction only applies to the color excess.  In this analysis we 
generally work with $B$ magnitudes, and therefore if the color excess
relation is purely due to dust then we expect $\beta = R_B = R_V + 1
\simeq 4$.  Interestingly, the current constraint on $\beta$ from
combined SNLS and low-z data is $\beta \sim 2 \pm 0.2$, which differs
from 4 by $> 10 \sigma$. A similar value was found by \citet{Tripp:98}.

What does $\beta \neq 4$ mean? If $\beta$ is interpreted as arising
from dust, then this would require $R_B = 2$, which is extreme.  On
the other hand, we know that dust {\it must} be present at some level.
One possible explanation is that the various color parameters are
actually measuring some combination of two effects: dust, as well as some
additional intrinsic color variations that are not related to $\Delta$
or $s$.  The $\beta$ for this additional piece is not known, but if it
is less than the value for dust, then the effective combined value will be
driven down.  This additional color parameter could be interpreted as
additional scatter in the measured colors of SN~Ia; most of the
fitters studied here (except SALT) do include terms for such scatter,
but assume that it has no effect on the overall luminosity
($\beta=0$).  A worry, then, is that the relative balance of dust and
the additional intrinsic color might vary between SNe in different
environments.  Alternatively, it is possible that the dust along the
line of sight to some SNe~Ia is different than standard dust, perhaps
due to scattering effects local to the SN \citep{Wang:05}.

\section{TESTING THE HUBBLE BUBBLE WITH OTHER FITTERS}
\label{sec:fits}
\nobreak 
The data sample used in the analysis of J07 incorporates a number
of SN with sparsely sampled light curves.  We have placed the
following additional requirements on our sample: First we require at
least one rest frame $B$ observation within $\pm 7$ days of the
estimated $B$ peak, in order to ensure that $m_B$ is
well measured, and at least one $U$ or $V$ observation within -7 to
+10 days for the color.  We exclude all SN with $s < 0.7$ or $s >
1.3$, since these are outside the model bounds for most of the
fitters. Finally, we require that $E\left(B-V\right)_{MW} < 0.4$ mag
as calculated from the dust maps of \citet{Schlegel:98} and $c < 0.6$
mag, both to avoid regions of anomalous dust.  This reduces the sample
of 95 SN used by J07 to 61 objects.  Since only MLCS2k2 is trained to
fit $I$ data, we remove this from all fits.  In order to check that
this reduction in the sample does not affect the results, we then
refit all of these SN using the most recent version of MLCS2k2.  We
find good agreement with J07, with $\deltah = 6.0\% \pm 1.9\%$ for
$cz_{\mbox{void}} = 7400$ km s$^{-1}$ and assuming a flat $\Omega_m=0.3$
$\Lambda$CDM universe.

\begin{figure}
\plotone{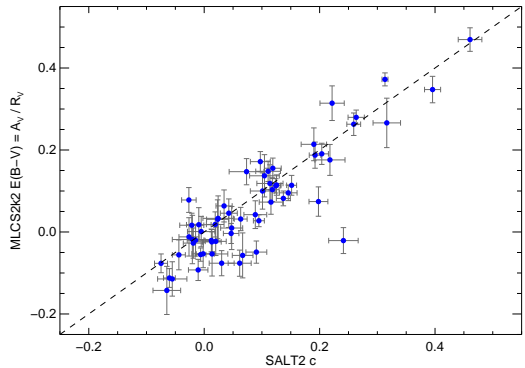}
\caption{SALT2 $c$ parameter compared with the equivalent MLCS2k2
 value, $A_V/R_V$ for our 61 SNe. The $A_V$ prior has been removed
 from the MLCS2k2 fits to allow easier comparison of the underlying
 fits.  Good agreement is also found in the times of maximum, peak
 magnitudes, and light-curve shape parameters (although for the latter
 the relations are non-linear).
\label{fig:salt2_mlcs_color} }
\end{figure}

\begin{figure}
\plotone{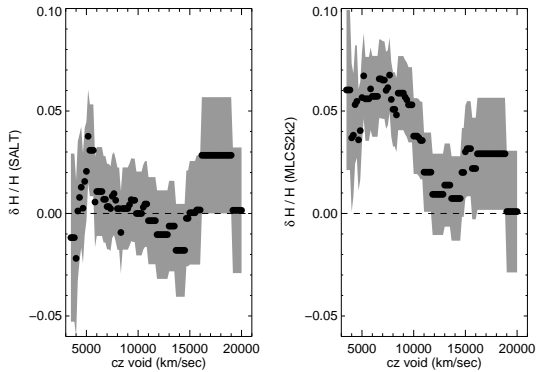}
\caption{Hubble bubble as a function of the velocity of the
 step in the Hubble constant, $cz_{\mbox{void}}$.  Left:
 Results for SALT (with $\beta=1.82$). Right: Results for MLCS2k2.
 The grey band is the error in \deltah .
 The constant value for $16000 < cz < 19000$ km s$^{-1}$ simply reflects the
 lack of any SN in this range.\label{fig:deltaH}
}
\end{figure}

We then fit the same photometry using the other light-curve 
packages\footnote{The raw light-curve parameters are available from
\url{http://qold.astro.utoronto.ca/conley/bubble}}.
The different packages show impressive agreement in their basic
derived light-curve parameters, as demonstrated in
figure~\ref{fig:salt2_mlcs_color}.  In order to measure \deltah with
SALT/SALT2/SiFTO, we must determine $\beta$.  This is calculated
from the SN data, but incorporating high-redshift SNLS data as well.
SALT, SALT2, and SiFTO give $\beta =$ 1.82, 1.75 and 2.31,
respectively, with errors of about $\pm 0.16$.  The different values
are not particularly important for the current analysis; all three can
be fixed to the mean value without qualitatively changing the results.
After applying the color and light-curve shape corrections, we obtain
$\deltah = 0.9\% \pm 2.0\%$, $-1.2\% \pm 2.1\%$, and $0.4\% \pm 2.1\%$
-- in other words, they do not support a Hubble bubble if $\beta$ is
fit to the data.  SALT and MLCS2k2 are compared in
figure~\ref{fig:deltaH} for various values of $cz_{\mbox{void}}$.  The
exact values are somewhat sensitive to the cuts applied to the sample.
However, it requires considerable hand tuning to find regions of
parameter space where any of the fitters besides MLCS2k2 detect a
Hubble bubble at greater than $1.4\, \sigma$, and in all cases MLCS2k2
gives much greater significance.

\begin{figure}
\plotone{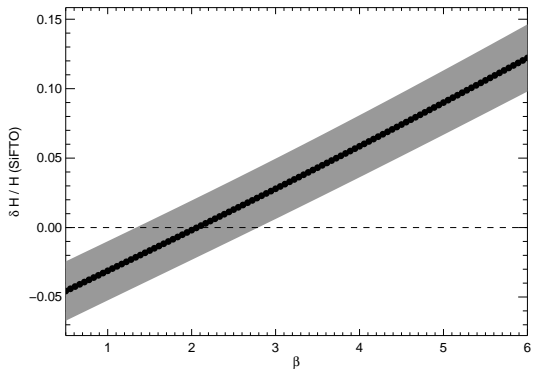}
\caption{\deltah\ vs.\ the value of $\beta$ used to convert the measured
 color parameter into a luminosity correction for SiFTO.  The
 relation is almost, but not quite, linear.  Fits to low-z and SNLS 
 data give $\beta = 2.35 \pm 0.16$. \label{fig:deltaHbeta} }
\end{figure}

One might be tempted to interpret this as an error in one or more of
the packages.  However, the actual cause is more interesting: if one
sets $\beta = 4.1$ for SALT/SALT2/SiFTO, they {\it do} find evidence
for a Hubble bubble at $> 2.5 \sigma$, as shown in
figure~\ref{fig:deltaHbeta}.  The color model for the most recent
version of $\Delta m_{15}$, which also favors the bubble, is similar
to that of MLCS2k2.  The question of the Hubble bubble therefore boils
down to the appropriate value of $\beta$ -- is it the $\sim 4$ of
Milky Way-like dust, or is something more complicated going on?

\begin{figure}
\plotone{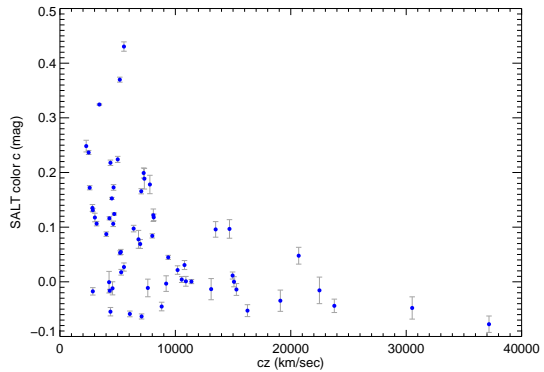}
\caption{Colors of the low-z supernova sample using SALT.
 The distant portion of the sample is bluer than the nearby
 portion, which is probably caused by selection effects. \label{fig:malmquist}}
\end{figure}

The value of $\beta$ is relevant because the nearby portion of the
low-z SN sample is redder than the distant portion
(figure~\ref{fig:malmquist}).  This is probably due to Malmquist bias
\citep{Malmquist:36} or other selection effects, since redder
SNe are fainter and harder to detect.  The effects of the color
correction are to make the blue SNe dimmer and the red SNe brighter
(relatively).  Malmquist bias has little effect as long as the
appropriate value of $\beta$ is applied across the whole sample;
however, if the right value is $\beta = 2$ and instead 4 is used, then
the distant portion will be made too faint, and the nearby portion too
bright, which is exactly the effect observed in J07.  MLCS2k2 usually
includes a prior on $A_V$, and the default version does not take into
account the redshift-dependent effects of Malmquist bias. Adjusting
the prior to reflect this increases the value of \deltah .

\section{THE VALUE OF $\beta$}
\nobreak 
\label{sec:beta}
If, in fact, $\beta = R_B$ (i.e., it is purely dust), then we should
note that a large range of $R_B$ values have been observed in different
environments. However, a sample mean value of $R_B = 2$ would be quite
surprising \citep{Draine:03}, even with selection effects.  Most
previous studies of large samples of SNe~Ia have found $R_B \simeq 3.5 \pm
0.3$ \citep{dellaValle:92, Riess:96, Phillips:99, Altavilla:04}.  This
value weakens the evidence for the Hubble bubble slightly to
$\sim 2\, \sigma$.

Our empirical values for $\beta$ were calculated in a model that
assumed a smooth local Hubble flow. It is possible that this could be
artificially suppressing \deltah , so we checked this by refitting
$\beta$ using only high-z SNLS SN, only the low-z SN below
$cz_{\mbox{void}}$, and simultaneously with our fits to \deltah.
These give essentially the same result ($\beta \sim 2$), albeit with
larger errors than the full sample.  Therefore, $\beta$ is robust
against the presence of a Hubble bubble.  Monte Carlo studies indicate
that the bias in $\beta$ ($b_{\beta}$) due to errors in the measurement
uncertainties and covariances is $\left| b_{\beta} \right| < 0.02$ for 
fairly extreme cases.

\begin{figure}
\plotone{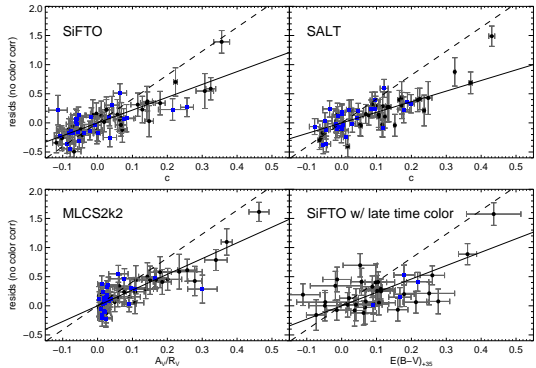}
\caption{Fits to $\beta$ for SiFTO (top left), SALT (top right),
 and MLCS2k2 (bottom left).  The residuals are compared to the best
 fitting Hubble line but without correction for the various color
 parameters, which are shown along the abscissas.  Bottom right: 
 Results for SiFTO using the late time colors as
 described in the text.  The solid lines are the best fit to $\beta$
 as given in the text. The dashed lines show $\beta = 4.1$, which is
 the expected value if the color excess relation is caused by
 Milky Way-like dust.  The black circles are for SNe with $cz < 7400$
 km s$^{-1}$, and the blue squares for $cz > 7400$ km s$^{-1}$.  
 The grouping around $A_V/R_V = 0$ for MLCS2k2 is the result of the $A_V$
 prior. \label{fig:betafits} }
\end{figure}

We can also analyze the results of the MLCS2k2 fits using the $\beta$
framework.  This differs from what is meant in J07 by fitting $R_V$,
which enforces a certain relation between the wavelength dependence
and scaling of the color excess relation.  We first remove the
extinction correction from the MLCS2k2 distance estimates, then
convert $A_V$ into $E\left(B-V\right)$ using $R_V$, and finally use
this to fit for the value of $\beta$ by minimizing the residuals with
respect for the Hubble line via a $\beta E\left(B-V\right)$ term.  We
work in $B$, and find $\beta = 2.7 \pm 0.3$, which again should be
compared with the expected value of 4.1.  Restricting the fit to only
SNe below $cz_{\mbox{void}}$ does not change the results.  The fits are
shown in figure~\ref{fig:betafits}.

Another technique for estimating the amount of extinction is to use
late time ($\sim 45$ days after peak) color measurements.  The idea is
that at late times all SNe~Ia have a simple relationship between intrinsic 
color and epoch, and this can be used to measure extinction \citep[][hereafter
P99]{Phillips:99}.  The evidence for this is based on a handful of SN
for which there is independent evidence for low extinction, and the
distribution of the measured late-time colors (J07, figure~6).  A
version of this is used in the training process of MLCS2k2.

This suggests one more test of $\beta$.  We take the subsample of our
61 SNe that have late-time color measurements (from J07) and use these
values as our color estimate to fit for $\beta$. Carrying out this
analysis out for all four fitters, we find $\beta \simeq 2.3 \pm 0.3$
for 37 SNe. In order to eliminate any ``cross-talk'' between the color
parameters and the peak magnitudes, we also tested this procedure
using only $B$ band data to fit the peak magnitude and light-curve
shape and obtained the same results.  Using the late-time colors
provided by P99 gives $\beta \simeq 1.5$ for 28 SNe. These results
again differ considerably from the expectation of 4.1.  

\section{CONCLUSIONS AND DISCUSSION}
\nobreak
We have demonstrated that the SN~Ia evidence for a Hubble bubble is
related to how SN colors are modeled.  All of the approaches
considered agree that the wavelength dependence of the color excess
relation is similar to that of dust, but disagree on whether or not
the relation between the measured color excess and the peak luminosity
is also dust-like.  Our fits give a value of $\beta \sim 2$, which if
interpreted via the CCM dust law, requires the extreme value $R_V \sim
1$.  Therefore, either a more complicated model of intrinsic SN
colors is required, which goes beyond a single light-curve shape-color
relation, or dust in the host galaxies of SNe~Ia is quite atypical of
Milky Way dust.  If the former, then the late-time colors of SNe~Ia
vary from SN to SN, and therefore do not provide a simple measure of
extinction.  If one does favor the single-parameter model with
Galactic dust, then the evidence for the Hubble bubble from SNe~Ia is
fairly strong.  These results depend on how accurately the light-curve
shape-intrinsic color relationship has been modeled, so it is
reassuring that the four fitters are in approximate agreement on the
value of $\beta$.

Requiring an additional intrinsic color relation beyond the
shape-color relation raises some challenges for SN research. Unless
this relation can be disentangled from dust, we must consider the
possibility that the balance of the two effects will change with
environment and redshift, perhaps even within the low-z sample, and
affect precision SN cosmology.  MLCS2k2 might require a more
sophisticated prior that takes into account the relation between
extinction, light-curve shape, intrinsic color, and would also have to
allow for any effects of this intrinsic color on SN luminosity.
SALT/SALT2/SiFTO might require different values of $\beta$ in different
environments or for different color thresholds.  Using an
inappropriate value of $\beta$ will mostly affect the most distant SN
in any survey, where Malmquist bias is important.  The potential
systematic for a given SN survey can be evaluated by multiplying the
uncertainty in $\beta$ by the change in color excess across the
sample. If the two-component color model is correct, then constraining
it will not be a trivial task, but it does hold out the possibility of
making SNe~Ia even better standard candles.

How can we resolve this issue?  A deeper nearby supernova sample that
has a similar color distribution at all distances out to $cz \sim
25000$ km s$^{-1}$ would provide a good test of the Hubble bubble, since an
incorrect value of $\beta$ would no longer introduce such an effect.
Note that SN cosmology analyses that restrict themselves to $z_{\rm
min} \gtrsim 0.015$ such as \citet{Astier:06, Riess:07, WoodVasey:07}
are not strongly affected by the existence of the bubble.  Determining
if a more complicated color model is necessary requires a different
approach.  A larger sample of SNe in low-extinction environments (like
elliptical galaxies), or at least in a narrow color range, could be
used to search for a non-dust-like color relation.  A wider baseline
of color measurements could also help this problem; it seems unlikely
that the wavelength dependence of the color excess relationship will
continue to look like dust at all wavelengths unless it really is
dust.  The SALT/SALT2 color excess relationship
\citep[figure~3]{Guy:07} displays tantalizing hints of departures from
the CCM law.  If these can be conclusively demonstrated, it would at
least prove that there is more going on than Milky Way-like dust, even
if it might not elucidate the underlying mechanism.

\acknowledgements
We thank Reynald Pain and Kathy Perrett for useful discussions.

\end{document}